\documentstyle[epsf,ijmpe1]{article}
\begin{document}

\runninghead{Varun Sheel et. al.}{Quark Propagator and Meson
Correlators in the QCD Vacuum}

\normalsize\textlineskip
\thispagestyle{empty}
\setcounter{page}{1}

\copyrightheading{Vol. 6, No. 2 (1997) 000--000}

\vspace*{0.88truein}

\fpage{1}
\centerline{\bf QUARK PROPAGATOR AND MESON CORRELATORS}
\vspace*{0.035truein}
\centerline{\bf IN THE QCD VACUUM}
\vspace*{0.37truein}
\centerline{\footnotesize VARUN SHEEL\footnote{Electronic address:
varun@prl.ernet.in}, HIRANMAYA MISHRA {\footnotesize and} 
JITENDRA C. PARIKH }
\vspace*{0.015truein}
\centerline{\footnotesize\it 
Theory Group, 
Physical Research Laboratory, Navrangpura,}
\baselineskip=10pt
\centerline{\footnotesize\it Ahmedabad 380 009, India }

\vspace*{0.225truein}
\publisher{(received date)}{(received date)}

\vspace*{0.21truein}
\abstracts{ Equal time, point to point correlation functions 
for spatially
separated meson currents are calculated with respect to a
variational construct  for the ground state of QCD.
 Given such an ansatz we make no further approximations
in the evaluation of the correlators.
Our calculations for the vector, axial vector and
scalar channels show qualitative
agreement with the phenomenological predictions, whereas
the pseudoscalar channel does not.  However, the
pseudoscalar correlator, when approximated by saturating
with intermediate one  pion states
agrees with results obtained from spectral density
functions parameterised by pion decay constant and
$<-\bar \psi \psi>$
value obtained from chiral perturbation theory.
We discuss this departure in the pseudoscalar channel , in
context of the quark propagation in the vacuum.}{}{}

\setcounter{footnote}{0}
\renewcommand{\thefootnote}{\alph{footnote}}

\vspace*{1pt}\textlineskip	
\section{Introduction}		
\vspace*{-0.5pt}
\noindent
Quantum Chromodynamics (QCD) in the low energy sector is 
nonperturbative and the vacuum structure here is 
nontrivial.\citeup{suryak} The vacuum structure of QCD has been 
studied since quite some time both with quark condensates 
associated with chiral symmetry breaking\citeup{nambu}  as well as 
with gluon condensates.\citeup{shut,hans} An interesting quantity 
to study with such a nontrivial structure of vacuum is the 
behaviour of current-current
correlators illustrating different physics involved at
different spatial distances. This has recently been emphasized
in a review by Shuryak\citeup{surcor} and studied through
lattice simulations.\citeup{neglecor} The basic point is that
the correlators can be used to study the interquark
interaction --- its dependence on distance. In fact 
they complement bound state hadron properties in the same way
that scattering phase shifts provide information about the
nucleon-nucleon force complementary to that provided by the
properties of the deuteron.\citeup{neglecor} 

We have recently considered the structure of QCD vacuum with
both quark and gluon condensates using a variational 
ansatz.\citeup{prliop} We shall use here such an explicit construct of
QCD vacuum obtained through energy minimisation\citeup{prliop}
to evaluate the meson correlators.

We organise the paper as follows. In section 2 we recapitulate 
the results of Ref.~7. We then discuss in section 3 the
quark propagation in our model of the QCD vacuum.  In section 4 we
define and calculate meson correlation functions. In section 5
we quote the results. Section 6 is devoted to the study of
the exceptional case of pseudoscalar correlator. Finally we
discuss the results in section 7.

We wish to make it clear  that in this work we have calculated 
correlations of two currents each having the quantum numbers of the
appropriate meson. The correlations are not between physical 
meson states.\cite{surcor}

\vspace*{1pt}\textlineskip	
\section{QCD Vacuum with Quark and Gluon Condensates} 
\vspace*{-0.5pt}
\noindent

We have considered the vacuum structure in QCD using a variational 
approach with both quark and gluon condensates.\citeup{prliop}
Here we shall very briefly recapitulate the results of the
same for the sake of completeness.
The trial variational ansatz for the QCD vacuum is taken as

\begin{equation}
|vac>=U_GU_F|0>,
\label{vacp}
\end{equation}
obtained through the unitary operators  $U_F$  and  $U_G$ for
quarks and gluons respectively on  the perturbative vacuum $|0>$. 

For the quark sector, the unitary operator $U_F$ is of the form   
 \begin{equation}
U_F
=\exp{({B_F}^\dagger-B_F)},
\label{unitop}
\end{equation}
 with the quark antiquark pair creation operator 
${B_F}^\dagger $ given by
 \begin{equation}
 {B_F}^{\dagger}=
\int \bigg[h(\vec k){{c^i}_{I}(\vec k)}^{\dagger}
(\vec \sigma \cdot \hat k)
{\tilde c}^i_{I}(-\vec k)
\bigg]\; d\vec k.
\label{bfbeta}
\end{equation}
\noindent
The operators $c^\dagger$ and $\tilde c$ create a quark and 
antiquark respectively when operating on the perturbative vacuum. 
They satisfy
the following quantum algebra in Coulomb gauge.\citeup{prliop} 
\begin{equation}
 [c^{i}_{Ir}(\vec k),c^{j}_{Is}(\vec k')^\dagger]_{+}=
 \delta _{rs}\delta^{ij}\delta(\vec k-\vec k')=
[\tilde c_{Ir}^{i}(\vec k),\tilde c_{Is}^{j}(\vec k')^\dagger]_{+}.
\end{equation}
Further $h(\vec k)$ is a trial function associated with 
quark antiquark condensates.

Clearly Eqs.\ (\ref{unitop}) and (\ref{bfbeta}) correspond 
to operator equations which create an arbitrary number of quark
antiquark 
pairs. In fact Eq.\ (\ref{bfbeta}) may be interpreted as an 
operator to create a Bose BCS state. We shall further assume 
h($\vec k$) to be spherically symmetric so that the non 
perturbative vacuum state will have the same symmetries as the 
perturbative vacuum i.e. zero momentum and angular momentum.

Similar considerations are applied for constructing the gluon
condensate function.Thus for the gluon sector we have,
 \begin{equation}
U_G
=\exp{({B_G}^{\dagger}-B_G)}
\: \; \; \; ; \quad
{{B_G}^\dagger}={1\over 2}
\int {f(\vec k){{{a^a}_{i}(\vec k)}^\dagger}
{{{a^a}_{i}(-\vec k)}^{\dagger}}d\vec k}.
\label{pairop}
\end{equation}
\noindent
Here $f(\vec k)$ is a trial function associated with gluon condensates
and ${a^a}_{i}(\vec k)$ the transverse gluon field  creation operators.

\noindent Clearly such a structure for the vacuum eventually 
reduces to a Bogoliubov transformation for the operators. 
One can then calculate the energy density functional given as

\begin{equation} 
\epsilon_{0} \equiv F(h(\vec k),f(\vec k)).
\end{equation}
The condensate functions $f(\vec k)$ and $h(\vec k)$ are to be
determined such that the energy density $\epsilon _0$ is a minimum. 
Since the functions cannot be determined analytically through
functional minimisation except for a few  simple cases,\citeup{grnv} 
we choose
the alternative approach of parameterising the condensate functions as
( with $k=|\vec k|$),

\begin{equation}
\tan 2h(\vec k)=\frac{A'}{(e^{R^2 k^2} - 1 )^{1/2}}.
\label{ansatz}
\end{equation}

\noindent
This corresponds to taking a Gaussian distribution for the
perturbative quarks in the nonperturbative vacuum.

Similarly, for the function $f(\vec k)$ describing the gluon
condensates we take the ansatz $\sinh f(\vec k)=Ae^{-Bk^2/2}$.

Further one could relate the quark condensate function to the
wave function of pion as a quark antiquark bound state\citeup{misra} 
and hence to the decay constant of pion.
It is clear from Eq. (\ref{ansatz}) and the ansatz for the
gluon condensate function
that even for $k=0$, $f(k)$ and $h(k)$ are non zero. 
Hence there is a finite probability to find low momentum states 
in the vacuum.

In Ref.~7 the energy density is minimised 
with respect to the condensate parameters
subjected to  the constraints that the pion decay constant
f$_\pi$ and the gluon condensate value 
$\frac{\alpha_s}{\pi}<G^a_{\mu\nu}{G^a}^{\mu\nu}>$
 of
Shifman Vainshtein and Zhakarov\citeup{svz} come out as the
experimental value of 93 MeV and 0.012 $GeV^4$ respectively. 

The results of such a minimisation showed the instability of
the perturbative vacuum to formation of quark antiquark as
well as gluon
condensates when the coupling became greater than 0.6. Further
the charge radius for the pion comes out correctly
($R_{ch}\simeq 0.65$ fm) for $\alpha_s =
1.28$. The corresponding values of A$^{'}$ and R of 
Eq.\ (\ref{ansatz}) are calculated to be
$A'_{min}\simeq 1$ and  $R\simeq 0.96$ fm.

It should be pointed out that local colour neutrality is
lost in the vacuum structure we have due to the generation of
mass like terms which is a limitation of our approach.

With the structure of QCD vacuum thus fixed from pionic
properties  and SVZ value we consider quark propagation
in the next section.

\vspace*{1pt}\textlineskip	
\section{Quark Propagation in the Vacuum}
\vspace*{-0.5pt}
\noindent

In the calculation of correlators, quark propagators enter in
a direct manner and hence it is instructive to study aspects
of the interacting propagator in some detail.\citeup{shuqprop} 
The reason for
doing this is two folds. We wish to know how it differs from a
free massive propagator i.e. how good it is to have a
``constituent quark'' picture and further to compare and
contrast with other approaches such as instanton liquid
model or vacuum dominance model based on operator product
expansion.

The equal time interacting quark Feynman propagator
in the condensate vacuum  is given as 
\begin{equation}
S_{\alpha \beta}(\vec x)   =  \left< \frac{1}{2} \left[
 \psi_{\alpha}^{i}(\vec x),
             \bar \psi_{\beta}^{i}(0) \right] \right>,
\end{equation}
In our model this reduces to
\begin{eqnarray}
S(\vec x) & = & \frac{1}{2} \frac{1}{(2 \pi)^3}
  \int e^{i\vec k .\vec x }d\vec k \left[ \sin 2 h(\vec k)
 - (\vec\gamma \cdot \hat k)~\cos 2 h(\vec k)\right]
\label{chiralprop1} \\
&  = &- \frac{i}{2 \pi^2} \frac{\vec\gamma \cdot \vec x}{x^4}
  +  \frac{1}{(2 \pi)^{3/2}} \frac{1}{2R^3} e^{-x^2 / (2R^2)}
 - \frac{i}{(2 \pi)^2} \frac{\vec\gamma \cdot \vec x}{x^2} I(x),
\label{chiralprop2}
\end{eqnarray}
where 
\begin{equation}
I(x) = \int_{0}^{\infty } \left( \cos k x - \frac{\sin k
x}{kx} \right)
   \frac{k e^{-R^2 k^2}}{1+(1-e^{-R^2 k^2})^{1/2}} dk,
\label{integral}
\end{equation}
\noindent
and $x=|\vec x|$, $k=|\vec k|$.

Clearly, the free massless propagator is given by 
$ S_{0}(x) = - \frac{i}{2 \pi^2} \frac{\vec\gamma \cdot \vec x}{x^4} $
which can be derived independently or from ( Eq.\ (\ref{chiralprop2})) 
in the limit $R \rightarrow \infty$ i.e. in the limit that the
condensate functions vanish.

It may be useful to note that the free massive propagator is
given as

\begin{equation}
S_0(m_q,\vec x)  = \frac{1}{(2 \pi)^2} \frac{m_q^2}{x} \left[
K_1(m_q x) - \; i \; \vec \gamma \cdot \hat x K_2(m_q x) \right]   ,
\label{masfree}
\end{equation}

\noindent where $K_1(m_q x)$ and $K_2(m_q x)$ are the first and second
order Bessel functions respectively.
 
\begin{figure}[t]
\vspace*{13pt}
\epsfxsize=5truein
\epsfbox[42 280 533 648]{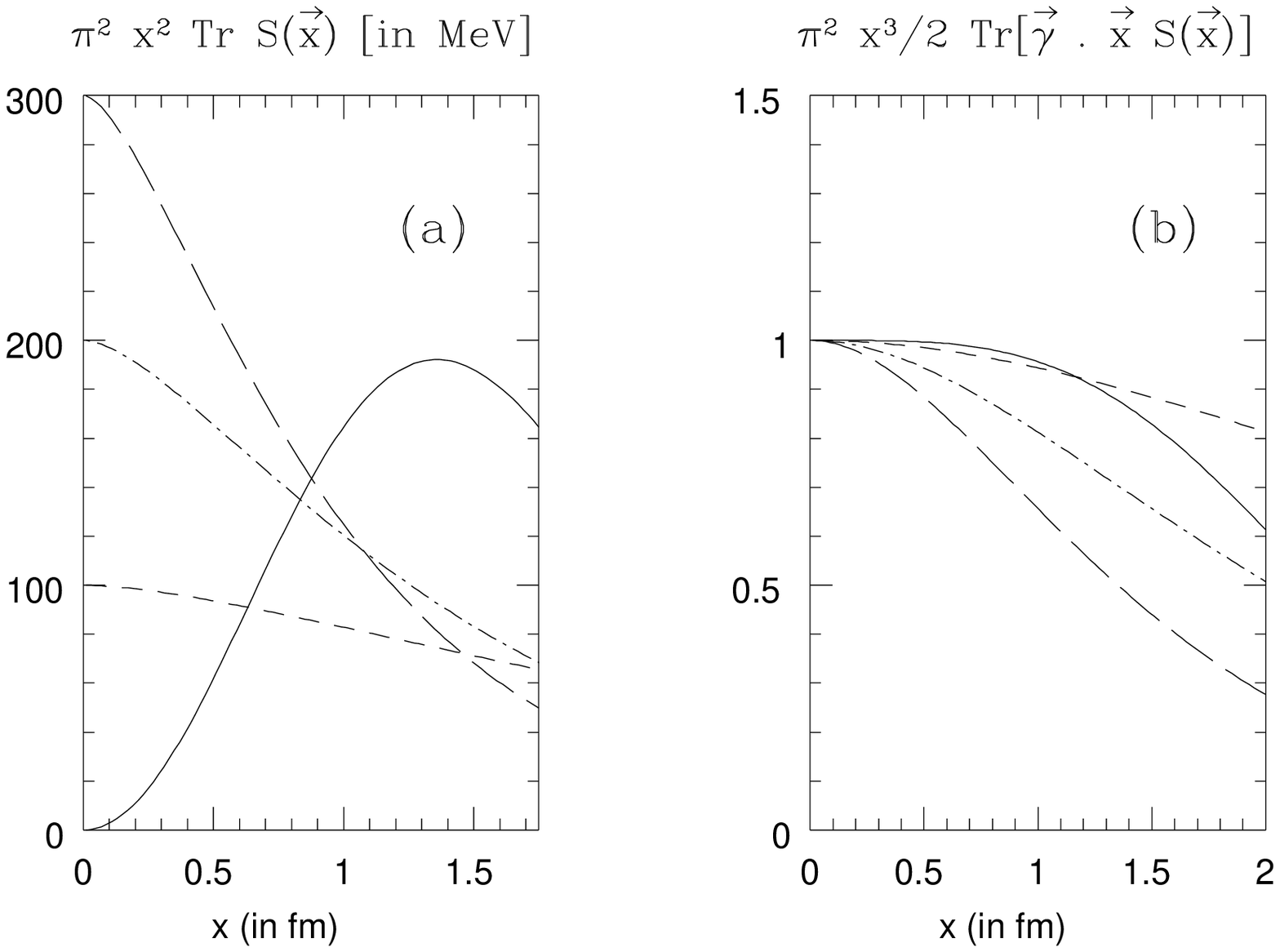}
\vspace*{13pt}
\fcaption{The two components of the quark propagator, Tr(S) (a)
and Tr[($\vec \gamma \cdot \hat x$)S] (b) versus the distance
x(in fm). The normalisation, indicated in the figure, have
been explained in the text. The solid line corresponds to
massless quark interacting propagator S(x). The three lines,
short dashed, dot-short dashed and long dashed correspond to a
massive free propagator with a mass of 100, 200 and 300 MeV,
respectively.
 \label{propfigure}}
\end{figure}

In Fig.~\ref{propfigure}
 we plot the two components Tr $ S(\vec x)$ and Tr $(\gamma
\cdot \hat x) S(\vec x) $ of the propagator for massless
interacting quarks given by Eq.\ (\ref{chiralprop2})
corresponding to the chirality flip and non-flip components 
considered by Shuryak and Verbaarschot.\citeup{shuqprop}
The first trace is normalised to the short distance limit of
the massive free quark propagator Tr~$ S_0(m_q,\vec x)/m_q$ which
from (Eq.\ (\ref{masfree})) is $1/(\pi^2 x^2)$. The
second trace is normalised to the free quark propagator which
is Tr $ (\gamma \cdot \hat x) S_0(x)= 2 \; i \; /(\pi^2 x^3)$.

To compare with the constituent quark models with an effective
constituent mass, we have also plotted the behaviour of free
massive quark propagator with masses 100 MeV, 200 MeV and
300MeV. In the chirality flip part, the propagator in the
condensate medium starts from zero, consistent with zero quark
mass at small distances, attains a maximum value of about 200
MeV at a distance of about 1.3 fm and then falls off gradually.
Further the interacting propagator overshoots the massive
propagators after about 0.8 fm. These features are
qualitatively similar to that of the instanton liquid model of
QCD vacuum considered in Ref.~11.

In the chirality non flip part, the interacting propagator
starts from 1, again consistent with the behaviour expected
from asymptotic freedom, but at larger separation it falls
rather slowly indicative of an effective mass of the order of
150~MeV. Again these features are similar to that of 
Ref.~11, though quantitatively there are
differences.

It may be amusing to consider the leading behaviour of the
propagator as $x \rightarrow 0$. In this limit, the first term
of Eq.\ (\ref{chiralprop1}) is given by

\begin{eqnarray}
T_1 & = &\frac{1}{2} \frac{1}{(2 \pi)^3} \int d\vec k  \sin 2 h(\vec k)
 + O(x^2) \\
  & = & \frac{1}{24} < - \bar \psi \psi > + O(x^2),
\end{eqnarray}
where $ < - \bar \psi \psi > = < \bar u u + \bar d d > $
for two flavours. Similarly the leading contribution from the
second term of Eq.\ (\ref{chiralprop1}) is
\begin{equation}
T_2 = - \frac{i}{2 \pi^2} \frac{\vec\gamma \cdot \vec x}{x^4}
+ O(x).
\end{equation}
Thus in the small x limit the interacting propagator of 
Eq.\ (\ref{chiralprop1}) reduces to

\begin{equation}
S(\vec x) = - \frac{i}{2 \pi^2} \frac{\vec\gamma \cdot \vec
x}{x^4}
  +  \frac{1}{24} <-\bar \psi\psi>  
\end{equation}
which is exactly the result of the vacuum dominance 
model\citeup{shuqprop} based on operator product expansion.

\vspace*{1pt}\textlineskip	
\section{Meson Correlation Functions}
\vspace*{-0.5pt}
\noindent

Consider a generic meson current of the form
\begin{equation}
J(x) = \bar \psi_{\alpha}^{i}(x) \Gamma_{\alpha \beta }  
\psi_{\beta}^{j}(x)
\label{current}
\end{equation}
\noindent where
$ x \; $is a four vector;
$\alpha$ and $\beta$ are spinor indices;
i and j are flavour indices;
$\Gamma$ is a $4 \times 4$ matrix $(1, \gamma_{5}, 
\gamma_{\mu}$  or $\gamma_{\mu} \gamma_{5} )$

Because of the homogeneity of the vacuum we define the conjugate
current to the above at the origin as,
\begin{equation}
\bar J(0) = \bar \psi_{\lambda}^{j}(0) \Gamma_{\lambda \delta }^{'}
\psi_{\delta}^{i}(0)
\label{currentbar}
\end{equation}
with $\Gamma^{'} = \gamma_{0} \Gamma^{\dagger } \gamma_{0} $

In general, the meson correlation function is defined as,
\begin{equation}
R(x) = < T J(x) \bar J(0) >_{vac}.
\label{cordef}
\end{equation}
From now on we assume that expectation values are always with
respect to the nonperturbative vacuum of our model,
hence we drop the subscript $vac$. 

Hence with Eqs.\ (\ref{current}), (\ref{currentbar}) 
and (\ref{cordef}) we have
\begin{equation}
R(x) = \Gamma_{\alpha \beta } \Gamma_{\lambda \delta }^{'} 
 < T \bar \psi_{\alpha}^{i}(x) 
\psi_{\beta}^{j}(x) \bar \psi_{\lambda}^{j}(0) 
\psi_{\delta}^{i}(0) >.
\label{cordef1}
\end{equation}
This reduces to the identity
\begin{equation}
R(x)  =  \Gamma_{\alpha \beta} \Gamma_{\lambda \delta}^{'} 
            < T \psi_{\beta}^{j}(x) \bar \psi_{\lambda}^{j}(0) > 
           <T \bar \psi_{\alpha}^{i}(x) \psi_{\delta}^{i}(0) > .
\label{cordef2}
\end{equation}
The above definition of $R(x)$ is exact since the four point
function does not contribute. 
In fact, in the evaluation of Eq.\ (\ref{cordef1}) we shall
have a sum of two terms. The first is
equivalent to the product of two point functions which is 
Eq.\ (\ref {cordef2}).
The second term arises from contraction of operators at the
same spatial point, related to
disconnected diagrams and thus can be discarded. 

In Eq.\ (\ref{cordef2}) the first term  can be identified 
as the interacting quark propagator
\[ S(x) = < T \psi^{j}(x) \bar \psi^{j}(0) > \]

It can be shown using the CPT invariance of the 
vacuum\citeup{bdrell} that the second term is given as 

\begin{eqnarray}
< T \bar \psi^{i}(x) \psi^{i}(0) > & = & - \gamma_{5} S(x) \gamma_{5}  
\\ \nonumber
      & = & - S(-x)
\end{eqnarray}
Hence the correlation function  of Eq.\ (\ref{cordef1}) becomes
\begin{equation}
R(x) = - Tr \left[ S(x) \Gamma^{'} S(-x) \Gamma \right].
\label{cordef3}
\end{equation}
Similarly the correlator for massless noninteracting quarks can
be given as

\begin{equation}
R_{0}(x)  =  - Tr \left[ S_{0}(x) \Gamma^{'} S_{0}(-x) \Gamma \right].
\label{cordef4}
\end{equation}

Our task is now to evaluate the expression (\ref{cordef3})
with the ansatz for QCD vacuum as given in Eq.\ (\ref{ansatz}).
Since we have evaluated the interacting quark propagator at
equal time we also consider the correlation functions 
(Eqs. (\ref{cordef3})-(\ref{cordef4})) at equal time.
This implies that the four vector $x$ in the above equations
reduces to the three vector $\vec x$

Having obtained the propagators in the earlier section, 
we can calculate the correlation
function, Eq.\ (\ref{cordef3}) for a generic current of the form
as in Eqs.\ (\ref{current}) and (\ref{currentbar}).

For convenience, we will consider the ratio of the physical
correlation function to that of massless noninteracting quarks
and obtain (with $x=|\vec x|$)
\begin{equation}
\frac{R(x)}{R_{0}(x)} = \left(1+\frac{1}{2} x^2 I(x) \right)^2
   + \frac{\pi}{8} \frac{x^6}{R^6}  e^{-x^2/R^2}
\frac{x^2 Tr\left[ \Gamma^{'} \Gamma \right]}{ x^{i} x^{j}
   Tr\left[\gamma^{i} \Gamma^{'} \gamma^{j} \Gamma \right]},
\label{gencor}
\end{equation}
\noindent
which is then evaluated in different channels with the
corresponding Dirac structure for the currents.

\begin{table}[h]
\tcaption{Meson currents and correlation functions \label{table1}}
\begin{tabular}{cccc}\\
\hline
CHANNEL & CURRENT & PARTICLE & CORRELATOR \footnote{The integral 
I(x) is defined in Eq.\ (\ref{integral})}\\
{} & {} & ($J^{P}$,MASS in MeV) &   
$\displaystyle \left[ \; \frac{R(x)}{R_{0}(x)} \; \right]$ \\
\hline
{} & {} & {} & \\
Pseudoscalar & $J^{p}=\bar u \gamma_{5} d$  & $ \pi^{0} (0^{-},135)$  
&  $\left[1+\frac{1}{2} x^2 I(x) \right]^2
   + \frac{\pi}{8} \frac{x^6}{R^6} e^{-x^2/R^2}$\\
{} & {} & {} & \\
\hline  
{} & {} & {} & \\
Scalar & $J^{s}=\bar u d$  & $ none (0^{+})$  
& $\left[1+\frac{1}{2} x^2 I(x) \right]^2
   - \frac{\pi}{8} \frac{x^6}{R^6} e^{-x^2/R^2}$  \\
{} & {} & {} & \\
\hline  
{} & {} & {} & \\
Vector & $J_{\mu}=\bar u \gamma_{\mu} d$  & $ \rho^{\pm} (1^{-},770)$  
& $\left[1+\frac{1}{2} x^2 I(x) \right]^2
   + \frac{\pi}{4} \frac{x^6}{R^6} e^{-x^2/R^2}$  \\
{} & {} & {} & \\
\hline  
{} & {} & {} & \\
Axial & $J_{\mu}^{5}=\bar u \gamma_{\mu} \gamma_{5} d$  
& $ A_{1} (1^{+},1100)$  
& $\left[1+\frac{1}{2} x^2 I(x) \right]^2
   - \frac{\pi}{4} \frac{x^6}{R^6} e^{-x^2/R^2}$  \\
{} & {} & {} & \\
\hline\\
\end{tabular}
\end{table}

\newpage

\vspace*{1pt}\textlineskip	
\section{Results}
\begin{figure}[hbp]
\vspace*{13pt}
\epsfxsize=5truein
\epsfbox[24 158 568 694]{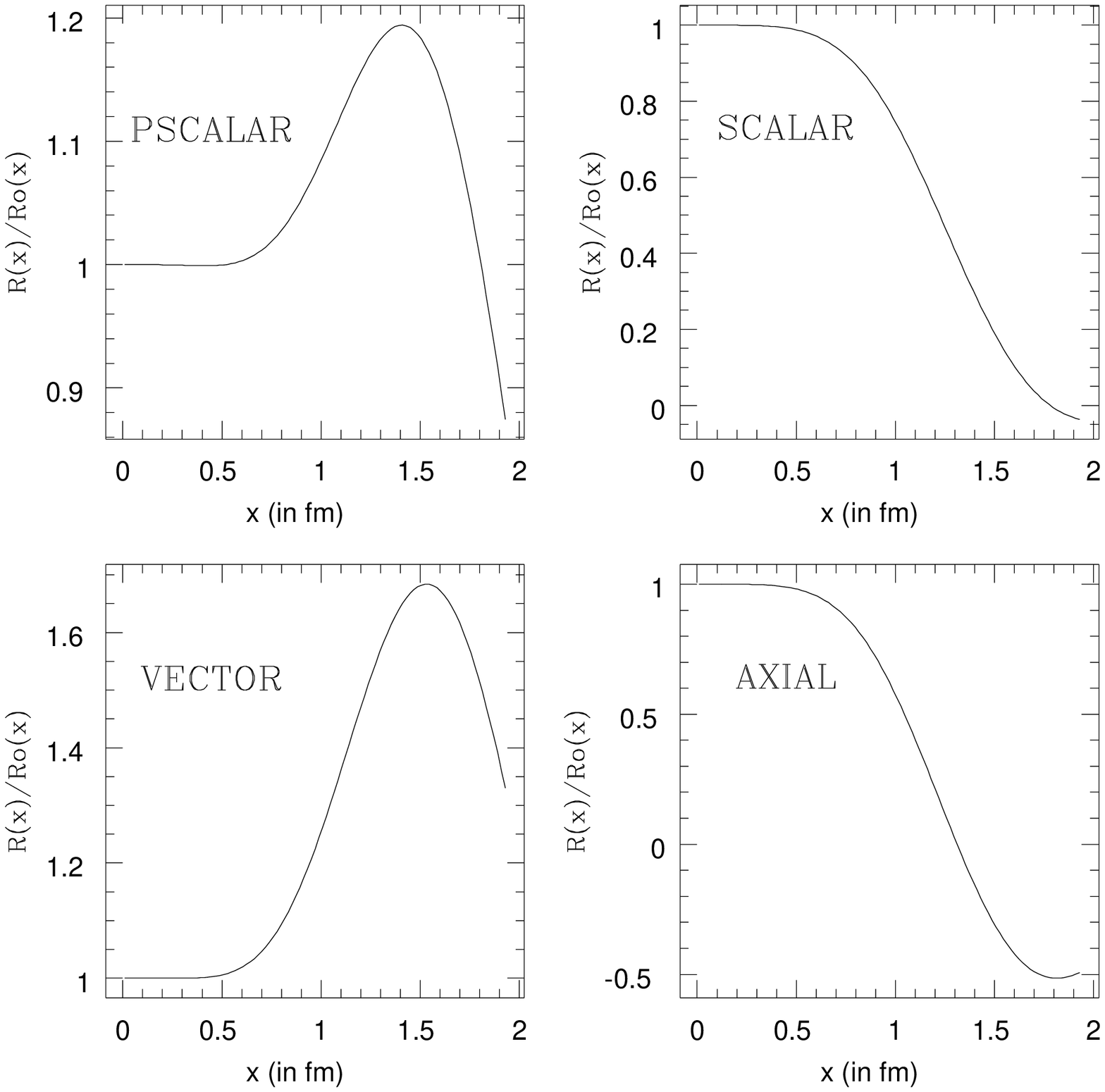}
\vspace*{13pt}
\fcaption{The ratio of the meson correlation functions in QCD
vacuum to the
correlation functions for noninteracting massless quarks,
$\displaystyle \frac{R(x)}{R_{0}(x)}  $,
Plotted vs. distance x (in fm) \label{exactfigure}}
\end{figure}

\vspace*{-0.5pt}
\noindent
We have studied the above ratio of correlators for four
channels. In each channel we associate the current with a
physical meson having quantum numbers identical to that of the
current. The results are shown in Table~\ref{table1} and in
Fig~\ref{exactfigure}.

We may notice some general features and relationships among
the correlators. The pseudoscalar correlator is always greater
than the scalar correlator and vector correlator is greater
than the axial vector correlator. We may emphasize here that
these relations are rather general in the sense that they do
not depend on the {\em explicit} form of the condensate
function and arise due to
the different Dirac structure of the currents which is reflected
in the generic expression for the correlation functions as in
Eq.\ (\ref{gencor}).
The behaviour of each channel is consistent with that predicted
by phenomenology except in the pseudoscalar case where the ratio
does not go as high as expected from phenomenology.  
We examine this in the next section.

\vspace*{1pt}\textlineskip	
\section{Pseudoscalar Channel}
\vspace*{-0.5pt}
\noindent
The explicit evaluation of the pseudoscalar correlator gives,
using Eq.\ (\ref{gencor}) 
\begin{equation}
\frac{R(x)}{R_{0}(x)} = 
  \left[1+\frac{1}{2} x^2 I(x) \right]^2
   + \frac{\pi}{8} \frac{x^6}{R^6} e^{-x^2/R^2}
\end{equation}
which may also be read off from column 4 of Table~\ref{table1}.
This is plotted as a function of $x$ in (Fig.~\ref{exactfigure}). 
As may be seen from (Fig.~\ref{exactfigure}) this ratio has a
maximum of $\sim$ 1.2 at $x \sim$ 1.3 fm. 
Phenomenologically\citeup{surcor} the peak is at 
$\sim$ 100 at $x \sim$ 0.5. In
order to compare our results with other calculations we
evaluate the same correlator approximately by saturating
intermediate states with one pion states.

With our definition of the correlation function 
(Eq.\ (\ref{cordef})) we have
\begin{eqnarray}
R(x) & = & < T J^p(x) \bar J^p(0) > \\
   & = & \frac{1}{2} \left( < J^p(\vec x) \bar J^p(0) >
         + < \bar J^p(0) J^p(\vec x) > \right) 
\end{eqnarray}
\noindent which comes from the definition of time ordered
product for equal time algebra. 

We now insert a complete set of intermediate states between
the two currents but retain only the one pion state
in the sum for the four point function. Thus,

\begin{eqnarray}
R(\vec x) & = &\frac{1}{2} \int (<vac \mid  J^p(\vec x)  \mid
        \pi^{a}(\vec p) > < \pi^{a}(\vec p) \mid \bar J^p(0)
\mid  vac> \nonumber \\ 
 & & + <vac \mid \bar J^p(0)  \mid \pi^{a}(\vec p) >
< \pi^{a}(\vec p) \mid J^p(\vec x) \mid vac >) d\vec p
\end{eqnarray}

Using translational invariance and the fact that for 
the pseudoscalar current
$J^{p}=\bar u \gamma_{5} d$ and $\bar J^p = -J^p $, 
the correlator may be written as

\begin{equation}
R(\vec x)  = \frac{1}{2} \int <vac \mid  J^p(0)  \mid
        \pi^{a}(\vec p) > < \pi^{a}(\vec p) \mid \bar J^p(0)
\mid  vac>
   \left(   e^{i \vec p . \vec x} + e^{- i \vec p . \vec x}
\right) d\vec p
\end{equation}

We may evaluate the above matrix element using the definition
of the pion decay constant given as\citeup{lee}
\begin{equation}
<vac \mid  J^{\mu a}_{5}(x)  \mid \pi^{a}(p) >
= \frac{i f_{\pi} p^{\mu}}{(2\pi)^{3/2} (2p_{0})^{1/2}}
e^{i p \cdot x}
\label{pdecay}
\end{equation}

\noindent where $ J^{\mu a}_{5} = [\bar \psi \gamma^{\mu} \gamma^5
\tau^a \psi ] $ is the axial current.

It can be shown\citeup{sakurai} that the divergence 
of the axial current gives the pseudoscalar current

\begin{equation}
\partial_{\mu} [\bar \psi \gamma^{\mu} \gamma^5 \tau^a \psi ] =
2 \; i \;m_q [\bar \psi \gamma^5 \tau^a \psi ] 
\label{dividentity}
\end{equation}
where $m_q$ is the current quark mass. Thus
taking divergence of both sides of Eq.\ (\ref{pdecay}) and using Eq.\ 
(\ref{dividentity}) we get,
\begin{equation}
2 \; m_q <vac \mid i J^{pa}(x)  \mid \pi^{a}(p) >
     =  \frac{- f_{\pi} m_{\pi}^{2}}{(2\pi)^{3/2} (2p_{0})^{1/2}}
          e^{i p \cdot x} 
           \label{divdecay}
\end{equation}

\noindent where we have used $ p^2 = m_{\pi}^2 $ .

In an earlier paper\citeup{misra} within our vacuum model  
and using the fact that pion is an approximate Goldstone mode
it was demonstrated that
saturating with pion states, gives the familiar current
algebra result

\begin{equation}
m^{2}_{\pi} = - \frac{m_q}{f_{\pi}^{2}} < -\bar \psi \psi >
\end{equation}
With this result we eliminate quark mass $m_q$ in Eq.\ (\ref{divdecay})
in favour of the
quark condensate to get the relation

\begin{equation}
<vac \mid J^{pa}(0)  \mid \pi^{a}(\vec p) >
=  \frac{i }{2 (2\pi)^{3/2} (2p_{0})^{1/2}} 
     \frac{< -\bar \psi \psi >}{f_{\pi}} 
\end{equation}

Using this, the expression for the pseudoscalar correlator now becomes
\begin{equation}
R(\vec x)  =  \frac{1}{64 \pi^3} \left( \frac{< -\bar \psi
\psi >}{f_{\pi}} \right)^2 \int \frac{1}{(p^2 + m_{\pi}^2 )^{1/2}}
\left(   e^{i \vec p . \vec x} + e^{- i \vec p . \vec x} \right)
d\vec p 
\end{equation}

The above integral can be evaluated using the standard 
integral\citeup{tabintprod}
\[
\int_{0}^{\infty} p(p^2 + \beta^2 )^{\nu - 1/2} sin(\alpha p) dp
= \frac{\beta}{\sqrt{\pi}} \left( \frac{2\beta}{\alpha} \right)^\nu
cos(\nu \pi) \Gamma (\nu + \frac{1}{2}) K_{\nu + 1} (\alpha \beta)
\]
for $\alpha > 0, Re \beta > 0$ and in the limit $\nu \rightarrow
0$. We then finally get  for the correlator 
(using saturation of pion states and with $x=|\vec x|$)

\begin{equation}
R(x) = \frac{1}{16 \pi^2} \left( \frac{< -\bar \psi
\psi>}{f_{\pi}} \right)^2 \frac{m_{\pi} K_1(m_{\pi}x)}{x}  
\end{equation}
The correlator for free massless quarks as calculated in the
earlier section for pseudoscalar is 

\begin{equation}
R_0(x) = \frac{1}{\pi^4 x^6}
\end{equation}

Hence the ratio is
\begin{equation}
\frac{R(x)}{R_{0}(x)} = \frac{\pi^2}{16} \left( \frac{< -\bar \psi
\psi>}{f_{\pi}} \right)^2  x^5 m_{\pi} K_1(m_{\pi}x)  
\end{equation}

\begin{figure}[t]
\vspace*{13pt}
\epsfxsize=5truein
\epsfbox[33 429 568 723]{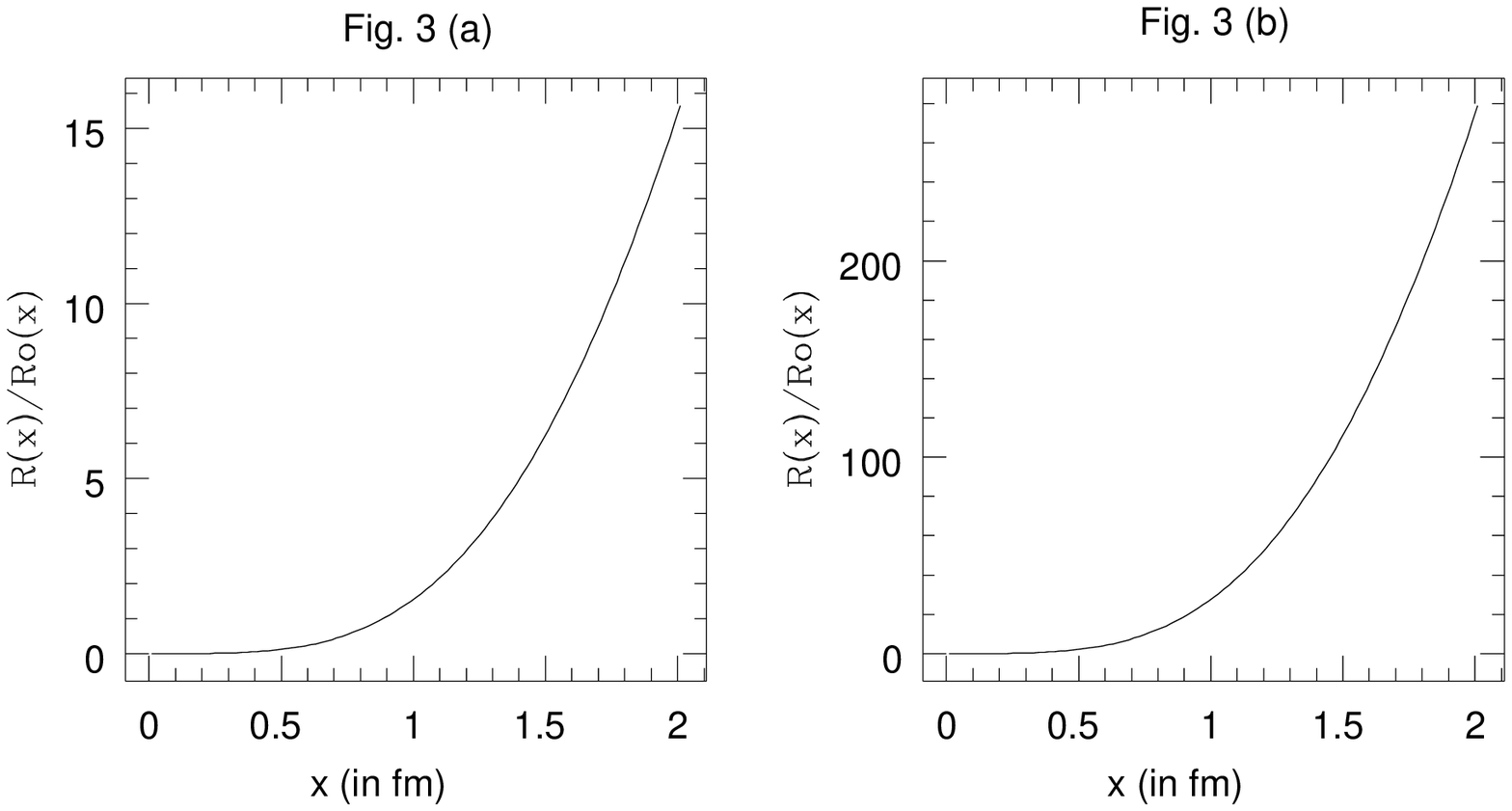}
\vspace*{13pt}
\fcaption{The pseudoscalar correlator plotted for  our value 
of $<-\bar \psi \psi> = $ (190 MeV)$^3$ in (a) and that of Shuryak
$<-\bar \psi \psi> = $ (307.4 MeV)$^3$  in (b)
 \label{satfigure}}
\end{figure}

We have plotted in Fig.~\ref{satfigure} this ratio for our 
value of $<-\bar \psi \psi>$ and that used by 
Shuryak.\citeup{surcor,surpap}
Note that our value of $<-\bar \psi \psi >$ is an output of the
variational calculation consistent with low energy hadronic
properties.\citeup{prliop}
 We thus observe that the
approximate calculation of the pseudoscalar correlator due to
saturation with one pion states (Fig.~\ref{satfigure}(a)) yields 
higher values ($\simeq$ 15 times more) as compared
to the calculations without saturation as an approximation 
(Fig.~\ref{exactfigure}). 
Thus the fermionic condensate model for QCD vacuum\citeup{prliop}
does not give as high values for the pseudoscalar correlator
as required by phenomenological results.
The value we have used for $< -\bar \psi \psi > \simeq$ (190 MeV)$^3$
 is smaller than Shuryak's value of (307.4 MeV)$^3$ \footnote{Our
 definition of the condensate value differs from the standard one by a
factor of $2^{1/2}$ } which appears in the parameterisation
of the physical spectral density through the coupling 
constant.\citeup{surpap} With his value of $< -\bar \psi \psi >$ the
ratio $R(x)/R_0(x)$ is shown in (Fig.~\ref{satfigure}(b))  
which agrees with phenomenology.

\vspace*{1pt}\textlineskip	
\section{Summary and Discussions}
\vspace*{-0.5pt}
\noindent
We have evaluated the mesonic correlators in this paper using a
variational construct for the QCD vacuum. Except for the
pseudoscalar channel the results show qualitative agreement
with phenomenological results.\citeup{surcor}
We have also shown that the quark propagation in our
construct for the QCD vacuum is almost identical to that in the
instanton model.

 Following Shuryak,\citeup{surcor,surpap} we also see 
that using current algebra
approach the pseudoscalar correlator rises sharply with
spatial separation. Let us recall that the current
algebra result also follows from the approximation of
saturating by one pion states in the normalisation of the pion
state.\citeup{misra}

It might appear that by suitably changing the value of $< -\bar \psi
 \psi >$ in our calculations without saturation one might be
able to reproduce all the phenomenological results. Actually
we find that it is not so. In fact, it adversely affects the 
correlators in the other channel which can be seen in
the expressions given in column 4 of  Table~\ref{table1}.
Also the results of our calculations are not very sensitive to
the parameter $R$.

In view of these findings, it is not clear whether saturation
of intermediate states by one pion states only in the
evaluation of the correlator, is sufficiently well justified.
We therefore think that a unified treatment of correlation
functions in all the channels is still not available. 

\nonumsection{Acknowledgements}
\noindent
VS and HM wish to acknowledge discussions with A. Mishra,
S. P. Misra and P. K. Panda.  We also thank A. Mishra for
a critical reading of the  manuscript.

\nonumsection{References}
\noindent

\def \sur { E.V. Shuryak, {\nineit The QCD vacuum,
 hadrons and the superdense matter} (World Scientific, 
Singapore, 1988).}

\def \nambu{ Y. Nambu, {\nineit Phys. Rev. Lett.} 
{\ninebf 4}, 380 (1960); Y. Nambu and G. Jona-Lasinio, 
{\nineit Phys. Rev.} {\ninebf 122},  345 (1961)
 ;ibid, 124  246 (1961); J.R. Finger and J.E. Mandula, 
{\nineit Nucl. Phys. } {\ninebf B199},  168 (1982);
 A. Amer, A. Le Yaouanc, L. Oliver, O. Pene and
 J.C. Raynal, {\nineit Phys. Rev. Lett. }{\ninebf 50},  87 (1983);
 ibid, {\nineit Phys. Rev. }{\ninebf D28},  1530 (1983); 
S.L. Adler and A.C.  Davis, {\nineit Nucl. Phys. }{\ninebf  B244},  
469 (1984); R. Alkofer and P. A.  Amundsen, {\nineit Nucl. Phys. }
{\ninebf B306},  305 (1988); A.C. Davis and A.M.  Matheson,
 {\nineit Nucl. Phys. }{\ninebf   B246},  203 (1984); 
S. Schramm and W. Greiner, 
{\nineit Int. Jour. Mod. Phys. }{\ninebf  E1},  73 (1992).}

\def\shut{D. Schutte, {\nineit Phys. Rev. }{\ninebf  D31}, 810 (1985).}

\def \hans {T. H. Hansson, K. Johnson, C. Peterson, 
           {\nineit Phys. Rev. }{\ninebf  D26},  2069 (1982).}

\def\surcor {E.V. Shuryak, {\nineit Rev. Mod. Phys. }
{\ninebf  65},  1 (1993)} 

\def \neglecor{M.-C. Chu, J. M. Grandy, S. Huang and 
J. W. Negele, {\nineit Phys. Rev. }{\ninebf  D48},  3340 (1993);
ibid, {\nineit Phys. Rev. }{\ninebf  D49},  6039 (1994).}

\def \prliop{ A. Mishra, H. Mishra, S.P. Misra, P.K. Panda and 
Varun Sheel, {\nineit Int. J. Mod. Phys. }{\ninebf  E5},  93 (1996).}

\def \svz{ M.A. Shifman, A.I. Vainshtein and V.I. Zakharov,
{\nineit Nucl. Phys. }{\ninebf  B147},  385 (1979), 448 and 519. }

\def \shuqprop{E. V. Shuryak and J. J. M. Verbaarschot,
{\nineit Nucl. Phys. }{\ninebf  B410},  37 (1993).}

\def \bdrell { See e.g. in J. D. Bjorken and S. D. Drell, 
{\nineit Relativistic Quantum Fields} (McGraw-Hill, New York, 1965)
p. 155 and 213.}

\def\mac {M. G. Mitchard, A. C. Davis and A. J. Macfarlane,
 {\nineit Nucl. Phys. }{\ninebf  B325},  470 (1989).} 

\def \lee { T. D. Lee,  
{\nineit Particle Physics and introduction to Field Theory} 
(Harwood Academic, 1982) p. 791. }

\def \sakurai { J. J. Sakurai,  
{\nineit Currents and Mesons} (Chicago Lectures in 
Physics, 1969) p. 18. }

\def \grnv { H. Mishra, S.P. Misra and A. Mishra,
{\nineit  Int. J. Mod. Phys. }{\ninebf  A3},  2331 (1988);
 S.P. Misra, {\nineit Phys. Rev. }{\ninebf  D35},  2607 (1987).}

 \def \misra{A. Mishra, H. Mishra and S. P. Misra, {\nineit Z. Phys. }
{\ninebf C57},  241 (1993) ; H.Mishra and S.P. Misra, 
{\nineit Phys Rev. }{\ninebf D48},  5376 (1993) ; A. Mishra and 
S.P. Misra, {\nineit Z. Phys. }{\ninebf C58},  325 (1993).}

 \def \tabintprod{{\nineit Table of Integrals, Series and 
Products} edited by I. S. Gradshteyn and I.M. Ryzhik 
(Academic Press,1980) p. 427.}

\def \shumeson{E. V. Shuryak and J. J. M. Verbaarschot,
{\nineit Nucl. Phys. }{\ninebf B410},  55 (1993).}

\def \shubar{T. Sch$\ddot{a}$fer, E. V. Shuryak and J. J. M.
Verbaarschot, {\nineit Nucl. Phys. }{\ninebf B412},  143 (1994).}

\def\surpap {E.V. Shuryak, {\nineit Nucl. Phys. }
{\ninebf B319},  541 (1989). }


\begin{thebibliography}{000}
\bibitem{suryak}\sur
\bibitem{nambu}\nambu
\bibitem{shut}\shut
\bibitem{hans}\hans
\bibitem{surcor}\surcor
\bibitem{neglecor}\neglecor
\bibitem{prliop}\prliop
\bibitem{grnv}\grnv
\bibitem{misra}\misra
\bibitem{svz}\svz
\bibitem{shuqprop}\shuqprop
\bibitem{bdrell}\bdrell
\bibitem{mac}\mac
\bibitem{lee}\lee
\bibitem{sakurai}\sakurai
\bibitem{tabintprod}\tabintprod
\bibitem{shumeson}\shumeson
\bibitem{shubar}\shubar
\bibitem{surpap}\surpap
\end{thebibliography}
\end{document}